\documentclass{article}

\usepackage[preprint]{neurips_2025}

\usepackage[utf8]{inputenc} % allow utf-8 input
\usepackage[T1]{fontenc}    % use 8-bit T1 fonts
\usepackage{hyperref}       % hyperlinks
\usepackage{url}            % simple URL typesetting
\usepackage{booktabs}       % professional-quality tables
\usepackage{amsfonts}       % blackboard math symbols
\usepackage{nicefrac}       % compact symbols for 1/2, etc.
\usepackage{microtype}      % microtypography
\usepackage{xcolor}         % colors
\usepackage{enumitem}
\usepackage{amsmath,amssymb}
\usepackage{threeparttable}
\usepackage{tabularx}
\usepackage{graphicx}
\usepackage{algorithm}
\usepackage{amsthm}
\usepackage{multirow}
\usepackage{makecell}
\usepackage{subfig}
\usepackage{wrapfig}

\theoremstyle{plain} 
\newtheorem{theorem}{Theorem}[section]
  
\theoremstyle{definition} 
\newtheorem{definition}{Definition}[section] 

\theoremstyle{remark}   
\newtheorem{remark}{Remark} 

\theoremstyle{assumption}

\title{Constrained Diffusers for Safe Planning and Control}

\author{%
  Jichen Zhang \\
  University of Oxford\\
  \texttt{jichen.zhang@worc.ox.ac.uk} \\
  \And
  Liqun Zhao \\
  University of Oxford\\
  \texttt{liqun.zhao@wolfson.ox.ac.uk} \\
  \And
  Antonis Papachristodoulou \\
  University of Oxford\\
  \texttt{antonis@eng.ox.ac.uk} \\
  \And
  Jack Umenberger \\
  University of Oxford\\
  \texttt{jack.umenberger@eng.ox.ac.uk} \\
}

\begin{document}

\maketitle

\begin{abstract}
  Diffusion models have shown remarkable potential in planning and control tasks due to their ability to represent multimodal distributions over actions and trajectories. However, ensuring safety under constraints remains a critical challenge for diffusion models. This paper proposes Constrained Diffusers, a novel framework that incorporates constraints into pre-trained diffusion models without retraining or architectural modifications. Inspired by constrained optimization, we apply a constrained Langevin sampling mechanism for the reverse diffusion process that jointly optimizes the trajectory and realizes constraint satisfaction through three iterative algorithms: projected method, primal-dual method and augmented Lagrangian approaches. In addition, we incorporate discrete control barrier functions as constraints for constrained diffusers to guarantee safety in online implementation. Experiments in Maze2D, locomotion, and pybullet ball running tasks demonstrate that our proposed methods achieve constraint satisfaction with less computation time, and are competitive to existing methods in environments with static and time-varying constraints. 
% Code is available in the supplemental material.
\end{abstract}

\section{Introduction}

In recent years, diffusion models -- powerful generative models that learn to model data distributions by gradually adding noise and then reversing the process -- have achieved remarkable success in the field of image generation \cite{sohl2015deep}. The forward diffusion process gradually corrupts expert data into noise through predefined Gaussian diffusion steps, while the reverse process learns to iteratively denoise this corruption using a trained neural network, ultimately reconstructing the original data distribution from random noise \cite{ho2020denoising}. This success has naturally been extended to policy representation in planning and control, outperforming traditional imitation learning methods, such as iteratively denoising trajectories for planning tasks \cite{janner2022planning} and modeling the policy as a return conditional diffusion model to obtain an optimal trajectory \cite{ajay2022conditional}. 

Despite these advances, applying diffusion model-based planning policies to real-world physical systems poses significant challenges, especially in terms of safety. In many practical scenarios, physical systems must operate within strict safety boundaries. Recent advances have explored incorporating constraints directly as conditions within generative models to achieve constraint satisfaction including classifier-guided\cite{dhariwal2021diffusion} and classifier-free\cite{ho2022classifier} based methods. Although these techniques have achieved remarkable success in image and text generation, their extension to planning and control faces significant challenges when constraints are nonconvex and time-varying, leading to training difficulties or the need to retrain for specific constraints \cite{janner2022planning, ajay2022conditional, ren2024diffusion, zheng2024safe}.

\begin{figure}
  \centering
  \begin{minipage}[b]{0.3\textwidth}
    \includegraphics[width=\textwidth]{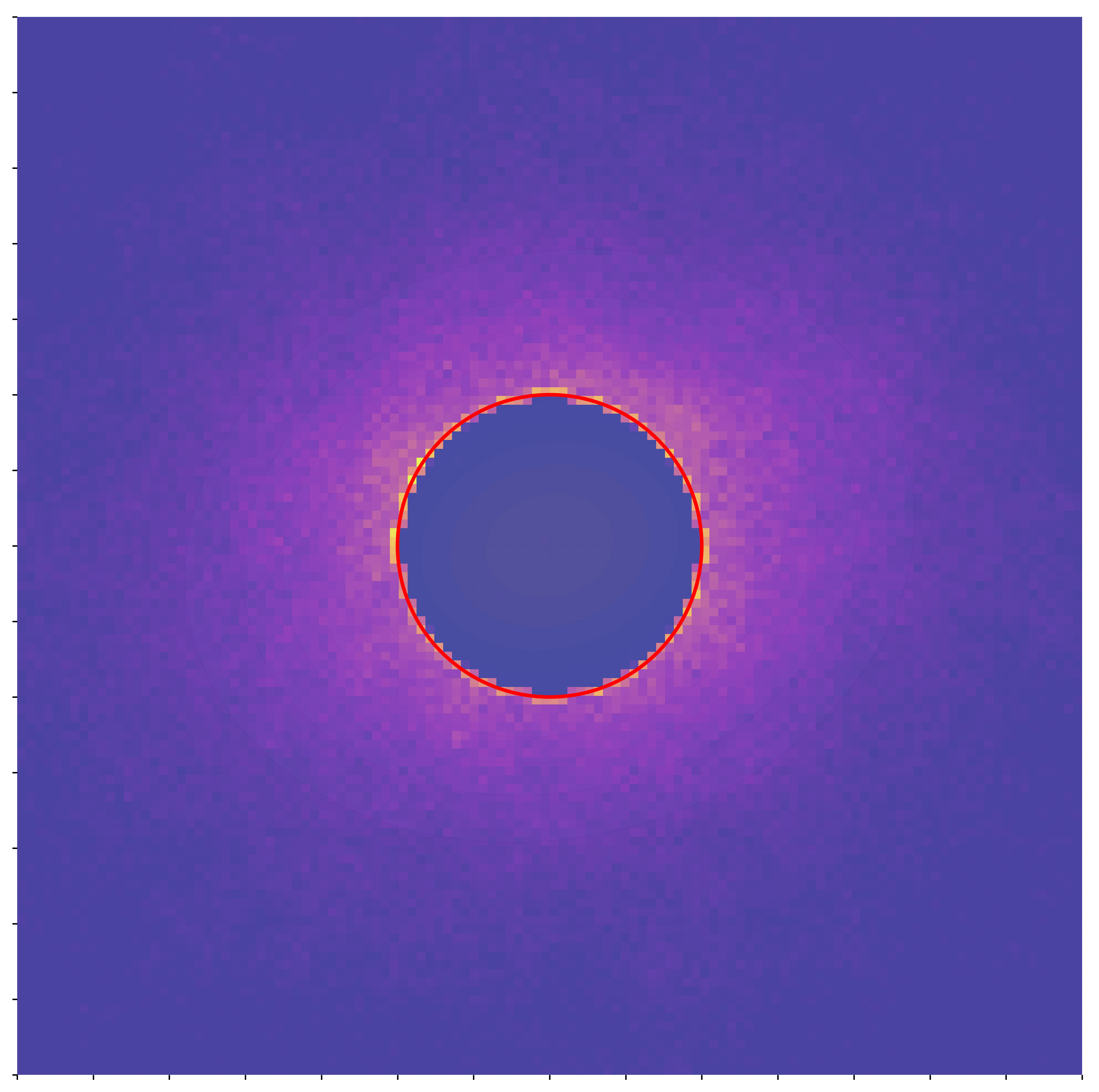}
    \centering
    \subfloat{(a) Projected Method} 
    \label{cs:a}
  \end{minipage}
  \hfill
  \begin{minipage}[b]{0.3\textwidth}
    \includegraphics[width=\textwidth]{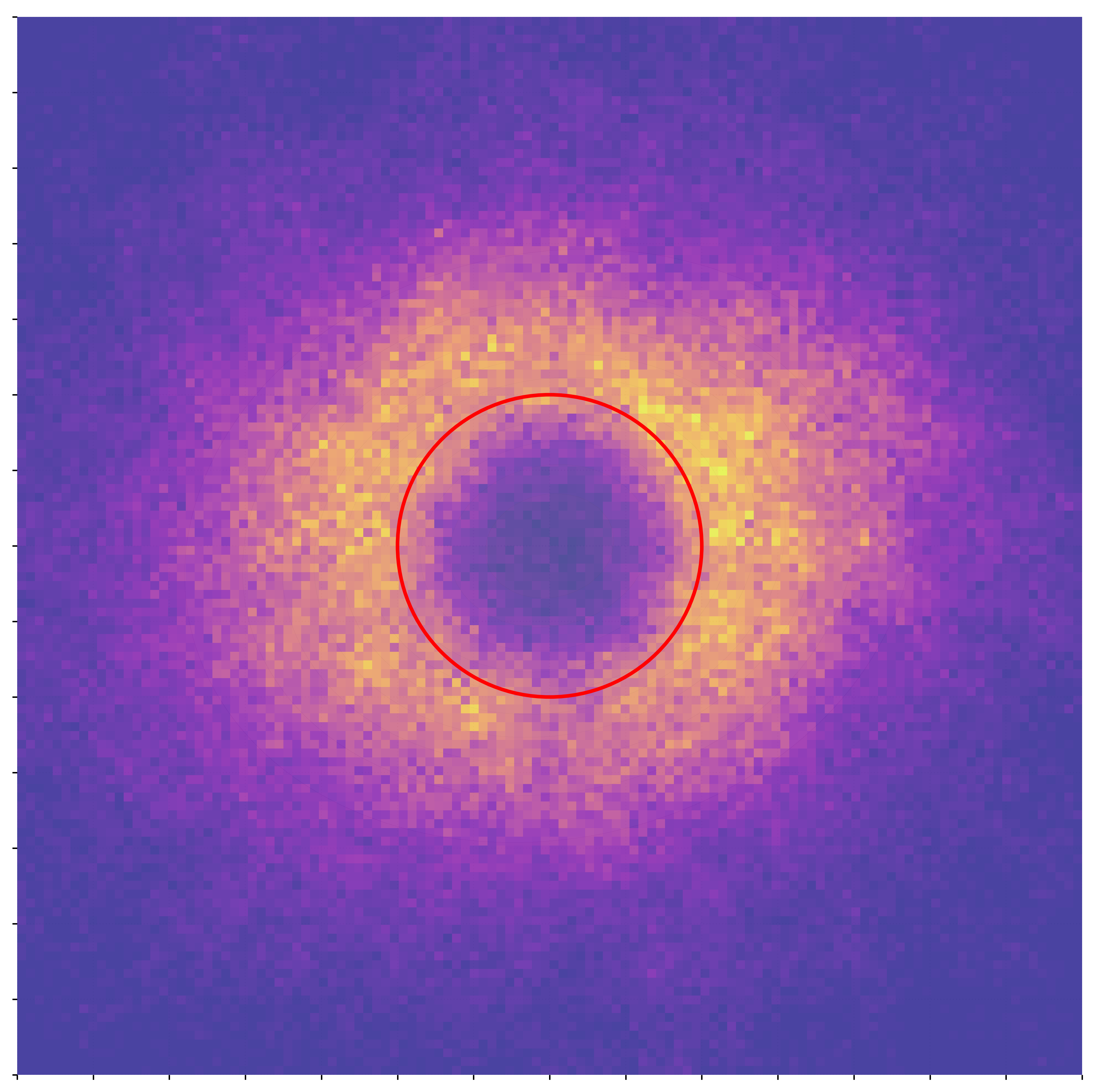}
    \centering
    \subfloat{(b) Primal-Dual Method}
    \label{cs:b}
  \end{minipage}
  \hfill
  \begin{minipage}[b]{0.3\textwidth}
    \includegraphics[width=\textwidth]{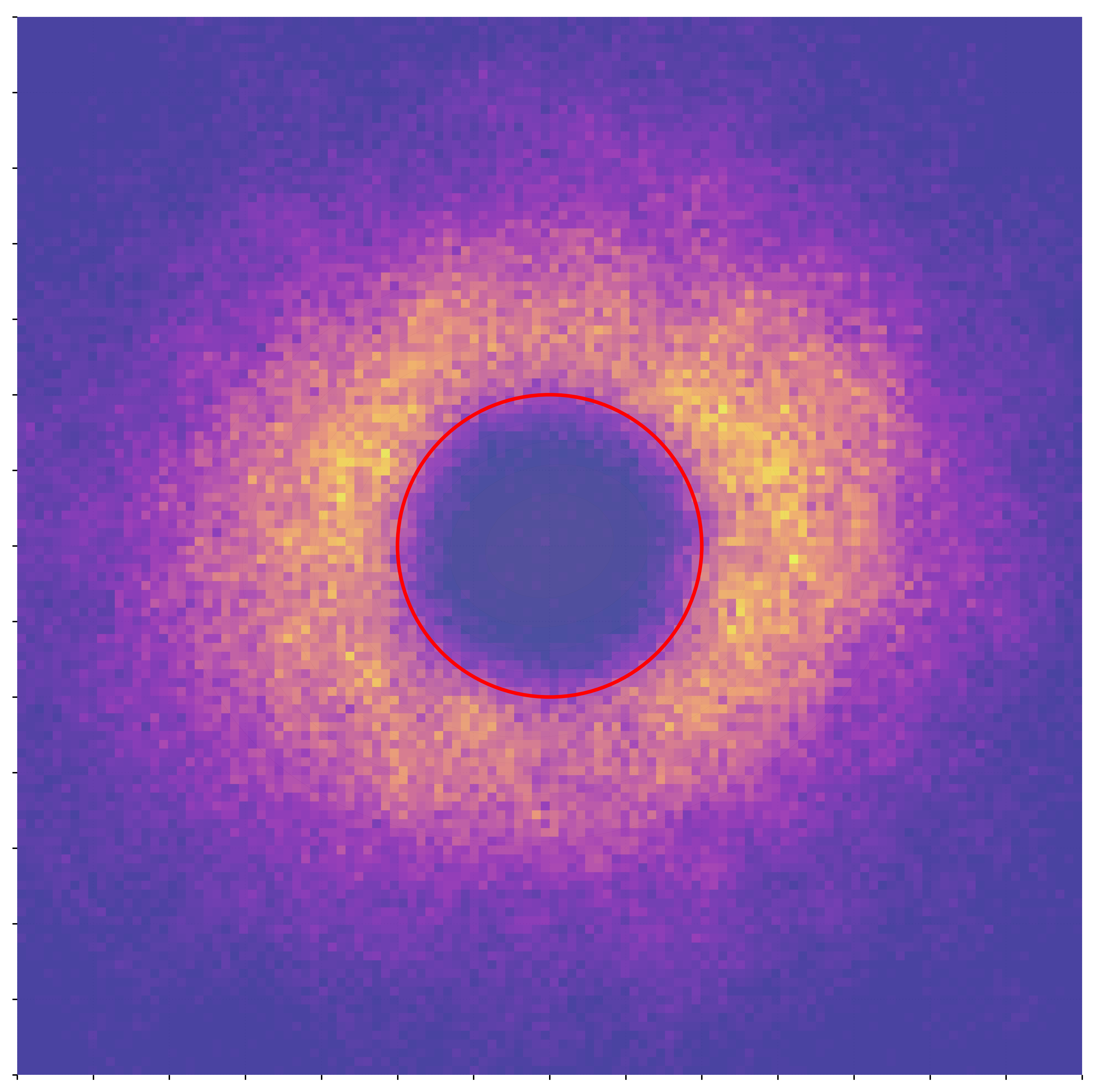}
    \centering
    \subfloat{(c) ALM}
    \label{cs:c}
  \end{minipage}
  
  \caption{Constrained sampling trajectories through (a) projected, (b) primal-dual and (c) augmented Lagrangian Methods. The goal is to use Langevin sampling methods to sample from a 2-dimensional Gaussian distribution with the constraint $x^2+y^2 \ge r^2$. Yellow regions show the high density of Langevin sampling trajectory distributions. Trajectories of projected method mostly concentrate on the constraint boundary while the other two are more distributed.}
  \label{cs}
\end{figure}

In this work, we propose Constrained Diffusers, formulating the problem from the perspective of Stochastic Gradient Langevin Dynamics \cite{song2019generative, wibisono2018sampling}. Specifically, we interpret the constrained diffusion generating problem as a constrained sampling problem, where the iterative solving process can be viewed as a stochastic gradient descent procedure. To incorporate constraints into the sampling process, we leverage techniques from constrained optimization, including projected, primal-dual, and augmented Lagrangian methods (ALM, Figure \ref{cs}). This integration allows the reverse diffusion steps to simultaneously optimize the generated trajectory while enforcing safety constraints. Building on the theoretical insights of \cite{karimi2024stochastic, bubeck2018sampling, chamon2024constrained, liu2021sampling, borkar2008stochastic}, we analyze the convergence of algorithms under mild conditions and provide guarantees for constraint satisfaction. Notably, our method enables direct constraint enforcement during the reverse diffusion process, eliminating the need for additional training when a pre-trained diffusion model is available for a given task. Furthermore, to ensure the system state stays within the safe set during online implementation, we use discrete control barrier functions as constraints in the constrained diffusion process to guide trajectory generation and employ an inverse dynamics model, trained by datasets, to derive actions feeding into environments.

% \begin{figure}
%   \centering
%   \includegraphics[width=\textwidth]{Figure_2.png}
%   \caption{Constrained sampling trajectories through projected, primal-dual and augmented Lagrangian Methods. The goal is to use Langevin sampling methods to sample from a 2-dimensional Gaussian distribution (as eclipse shown) with the constraint $x^2+y^2 \ge r^2$. Green points are the start points, red lines show the Langevin Sampling trajectories and blue points are the final points after certain sampling steps.}
%   \label{cs}
% \end{figure}

The main contributions of our work are summarized as follows:

\begin{itemize}[leftmargin=2em]
\item  We propose Constrained Diffusers, a novel framework that enforces constraints on trajectories generated by pre-trained diffusion models, modifying reverse diffusion process through integrating constrained sampling methods: Projected, Primal-Dual and Augmented Lagrangian methods, avoiding costly retraining or model modifications during deployment.
\item We incorporate discrete control barrier functions as constraints for constrained diffusers to guide trajectory generation, and then use an inverse dynamics model based on the constraint-satisfied trajectories to obtain actions, ensuring that the state remains within the safe set during the online implementation stage.
% \item We analyze the convergence of our proposed iterative algorithms under mild conditions, ensuring their stability and effectiveness when integrated into diffusion sampling, providing a rigorous theoretical foundation for enforcing constraints in the generating process.
\item We demonstrate the effectiveness of our proposed methods in Maze2D, locomotion, and pybullet ball running tasks, showing that our method effectively enforces constraints with competitive performance and computational efficiency.
\end{itemize}

\section{Background}
\label{gen_inst}

\subsection{Diffusion Models and Langevin Dynamics}
\textbf{Diffusion models} is a type of generative model inspired by principles from thermodynamics and works by simulating a two-step process. The forward process perturbs data $x_0 \sim q(x_0)$ by gradually adding noise over time $t \in [0, T]$,  and the reverse process reconstruct the target data distribution from a simple prior distribution, which can be formulated as the following SDEs \cite{song2020score}:
\begin{equation}
    dx_t = f(x_t, t) dt + g(t) dW_t, \quad dx_t = [f(x_t, t) - g(t)^2 \nabla_{x_t} \log p_t(x_t)] dt + g(t) d\bar{W}_t,
\end{equation}
where $W_t$ and $\bar{W}_t$ are standard Wiener processes running forward and backward in time, $f(x_t, t)$ and $g(t)$ are drift and diffusion coefficients chosen such that $p_T(x_T)$ approximates a simple prior, e.g., $N(0, I)$, and $\nabla_{x_t} \log p_t(x_t)$ is the score function of the marginal distribution $p_t(x_t)$.

\textbf{Langevin Dynamics} is a method that leverages gradient information and noise to sample from a given distribution, resembling simulating the diffusion reverse SDE \citep{song2020score} starting from $x_T \sim p_T$ and evolving backward to $t=0$. Standard Langevin Monte Carlo (LMC) aims to sample from a target distribution $p(x) \propto e^{-f(x)}$ using the update rule derived from the Langevin SDE:
\begin{equation} \label{eq:langevin_sde}
    dx(t) = -\nabla f(x(t)) dt + \sqrt{2} dW(t).
\end{equation}
In practice, due to the challenges in directly computing the path of SDE, we often use a discrete-time approximation, Stochastic Gradient Langevin Dynamics (SGLD):
\begin{equation}
    x_{t+1} = x_t + \frac{\epsilon}{2} \nabla_x \log p(x_t) + \sqrt{\epsilon} z_t
    \label{eq:ula}
\end{equation}
where $\epsilon$ is the step size, and $z_t \sim \mathcal{N}(0, I)$ is standard Gaussian noise. By iterating this update rule with an appropriate step size $\epsilon$, the sequence $x_t$ can be considered as samples drawn from $p(x)$. 

\textbf{Denoising Diffusion Probabilistic Model (DDPM)} can be connected to a Langevin-like update rule through its iterative denoising process. In DDPM, the forward process gradually adds Gaussian noise to data over $T$ steps, defined by $q(x_t|x_{t-1}) = \mathcal{N}(x_t; \sqrt{1-\beta_t}x_{t-1}, \beta_t\mathbf{I})$, where $\beta_t$ controls the noise schedule. The reverse process, $p_\theta(x_{t-1}|x_t)$, approximates the data distribution by learning the noise to denoise via a neural network, enabling the following reverse update:
\begin{equation}
x_{t-1} = \frac{1}{\sqrt{1-\beta_t}}\left(x_t - \frac{\beta_t}{\sqrt{1-\bar{\alpha}_t}}\epsilon_\theta(x_t,t)\right) + \sqrt{\beta_t}z,
\label{eq:6}
\end{equation}
where $\epsilon_\theta$ is the learned noise estimate and $\bar{\alpha}_t = \prod_{s=1}^t (1-\beta_s)$. This reverse step resembles Langevin dynamics, where the score function $\nabla_{x_t}\log p(x_t)$ is implicitly approximated by the denoising model's prediction of the noise component.  According to \cite{luo2022understanding}, we can rewrite the score function with respect to the noise term by combining Tweedie’s Formula with this reparameterization:
\begin{equation}
\nabla_{x_t} \log p(x_t) = -\frac{\epsilon_\theta(x_t,t)}{\sqrt{1-\bar{\alpha}_t}} = -\frac{x_t - \sqrt{\bar{\alpha}_t}x_0}{(1-\bar{\alpha}_t)}.
\label{eq:7}
\end{equation}
This reformulates the DDPM reverse process update into a Langevin Sampling process in terms of the score function $\nabla_{x_t} \log p(x_t)$ instead of the noise estimate $\epsilon_\theta(x_t,t)$. 
\begin{equation}
x_{t-1} = x_t + \frac{\beta_t}{2} \nabla_{x_t} \log p(x_t) + \sqrt{\beta_t}z.
\label{eq:8}
\end{equation}
% Since$ \frac{1}{\sqrt{1 - \beta_{t}}} \approx 1 + \frac{\beta_{t}}{2}
% $ for small $\beta_t$, we have the following update:
% \begin{equation}
% x_{t-1} = x_t +\beta_t \nabla_{x_t} \log p(x_t)+ \frac{1}{2}x_t + \sqrt{\beta_t}z, \quad z \sim \mathcal{N}(0,\mathbf{I}).
% \label{eq:9}
% \end{equation}
% To simplify the analysis and align with a Langevin-like form, we assume that the score function dominates the update (We will analyze our algorithm without the assumption in Appendix). Therefore, the final Langevin-like update, incorporating the score function, is:
% \begin{equation}
% x_{t-1} = x_t +\beta_t \nabla_{x_t} \log p(x_t) + \sqrt{\beta_t}z, \quad z \sim \mathcal{N}(0,\mathbf{I}).
% \label{eq:10}
% \end{equation}
This formulation bridges the DDPM reverse process to stochastic gradient-based Langevin dynamics, aligning denoising with probabilistic inference, forming the basis for leveraging optimization techniques within the sampling process. To ensure consistency in the time scale, we assume the iterative process proceeds from $T$ to 0 as $T \to \infty$ in this paper.

\subsection{Constrained Sampling}
In practical applications, it is crucial to generate samples that satisfy specific constraints or physical laws. This motivates the problem of \textit{constrained sampling}, where the goal is to sample from a distribution $q$ that is "close" to a reference distribution $p$ (e.g., the distribution implicitly defined by expert data) while satisfying certain constraints. As formulated by \cite{chamon2024constrained}, this can be posed as an optimization problem in the space of probability measures:
\begin{equation} \label{eq:constrained_sampling_opt}
    q^* = \mathop{\arg\min}_{q \in \mathcal{P}_2(\mathcal{R}^d)} \quad \mathrm{KL}(q || p) \quad \text{s.t.} \quad \mathbb{E}_{x \sim q}[g(x)] \le 0,
\end{equation}
where $g(\cdot)$ represent inequality constraint functions, respectively, and $\mathrm{KL}(q || p)$ measures the divergence from the reference distribution $p$.

For the task we aim to solve, we do not assume constraints are convex. Therefore, our approach involves adapting existing methods and extending them to non-convex settings, while proving the effectiveness of the proposed algorithm in experiments and giving theoretical analysis.
\subsection{Discrete Control Barrier Functions}
To enforce safety constraints in dynamical systems, Control Barrier Functions (CBFs) are designed to ensure forward invariance of a desired safe set $\mathcal{C}$, which is typically defined as the superlevel set of a continuously differentiable function $h(x)$, i.e., $\mathcal{C} = \{x \in \mathbb{R}^d : h(x) \ge 0\}$ \cite{ames2019control}. 

% For a general continuous-time deterministic system $\dot{x} = f_{dyn}(x,u)$ where $u$ is a control input, a function $h(x)$ is a CBF if there exists an extended class $\mathcal{K}$ function $\alpha$ such that for all $x \in \mathcal{C}$, 
% \begin{equation}
% \label{cbf}
%     \sup_{u \in \mathcal{U}} [ \frac{\partial h}{\partial x} f_d(x,u) ] \ge  - \alpha(h(x)),
% \end{equation}
% where $\mathcal{U}$ is the set of admissible control inputs. This condition ensures that for any state $x$, a control input $u$ can be found that makes $\dot{h}(x,u) \ge -\alpha(h(x))$, preventing the state from leaving the safe set $\mathcal{C}$. 
For a general discrete-time deterministic system $x^{\tau+1} = f_d(x^\tau, u^\tau)$, where $x^\tau \in \mathbb{R}^d$ is the system state at time step $\tau$, $u_\tau \in \mathcal{U} \subset \mathbb{R}^m$ is the control input, and $f_d$ denotes the discrete-time system dynamics, a function $h(x)$ is a Discrete Control Barrier Function (DCBF) if there exists an extended class $\mathcal{K}$ function $\alpha$ such that for all $x \in \mathcal{C}$\cite{agrawal2017discrete}:
\begin{equation}
\label{dcbf}
    h(f_d(x^\tau, u^\tau)) - h(x^\tau) \ge -\alpha(h(x^\tau)).
\end{equation}
This condition ensures that the decrease in $h(x)$ between discrete time steps is bounded, thus preserving the forward invariance of the safe set $\mathcal{C}$. In practice, this often involves solving a quadratic programme (QP) at each time step to find a control input $u$ that satisfies the CBF condition while minimally deviating from a nominal control policy.

% This approach offers a powerful mechanism for enforcing hard, state-wise constraints directly within the generative or sampling dynamics. It is particularly relevant when the generated samples must strictly adhere to physical limits, operational boundaries, or other invariant conditions throughout their evolution. While sharing the spirit of barrier methods like those mentioned in \cite{kook2022sampling}, CBFs provide a formal framework from control theory for synthesizing these corrective actions, complementing other constrained sampling techniques by offering a continuous-time perspective on constraint adherence.

\section{Problem Statement}

\begin{figure}
  \centering
  \includegraphics[width=\textwidth]{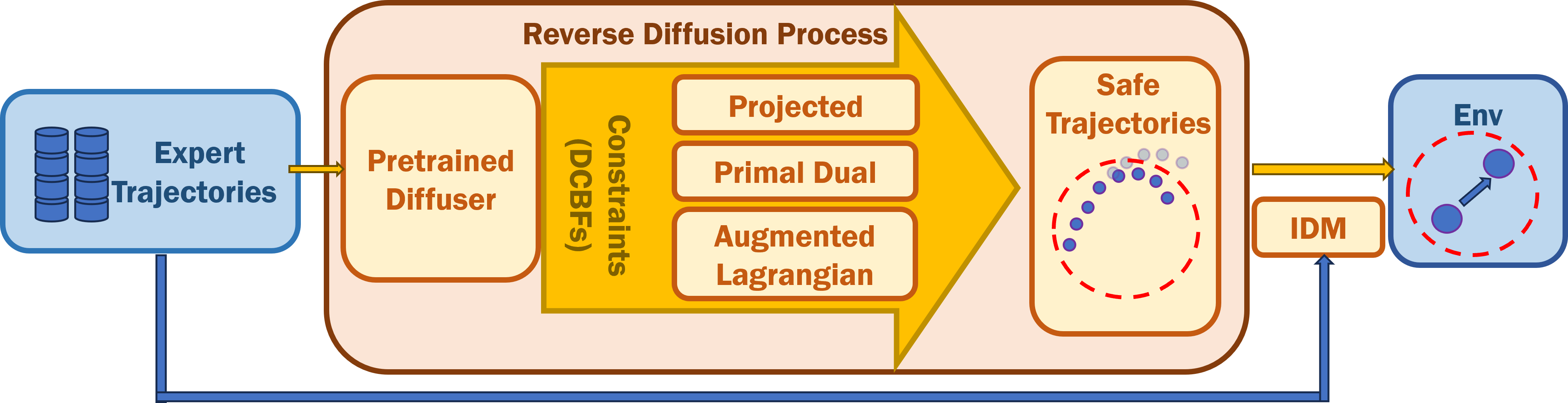}
  \caption{Overall Framework. We use expert trajectories to train the diffuser, apply constrained sampling methods for DCBFs in the reverse sampling process to get safe trajectories and finally use an inverse dynamics model to obtain actions, ensuring the safety in online implementation.}
  \label{frame}
\end{figure}

The generative power of diffusion models extends to sequential decision-making problems, leveraging the models to represent distributions over trajectories or action sequences. Specifically, these models are trained to model the distribution over expert demonstrations by corrupting trajectories $(x^0,\ldots, x^\mathcal{T})$ with noise in a forward diffusion process and learning to reverse this corruption during training\cite{janner2022planning}. 
% This framework aligns naturally with the goals of imitation learning, where the objective is to recover a policy that reproduces expert-like behavior without access to explicit reward signals. Diffusion models, due to their generative flexibility, can capture multimodal behavior and produce diverse trajectories that go beyond deterministic policy cloning, offering an alternative to traditional imitation learning approaches.
Despite the demonstrated success of diffusion models in trajectory planning and control, their practical deployment in safety-critical systems faces a fundamental challenge which is enforcing dynamic constraints during sampling without costly retraining or architectural modifications.

To address this, we consider the following problem: given a pretrained diffusion model defining an implicit distribution $p(x^{0:\mathcal{T}})$ over trajectories, and functions $g(x^{0:\mathcal{T}})$, design an alternative reverse process that produces a "similar" trajectory distributions satisfying
\begin{equation}
\label{problem}
q^* = \mathop{\arg\min}_{q \in \mathcal{P}_2(\mathcal{R}^d)} \quad \mathrm{KL}(q || p) \quad s.t. \quad
  \mathbb{E}_{x^{0:\mathcal{T}} \sim q}[g(x^{0:\mathcal{T}})] \le 0,\quad \forall\, \tau=0,\dots,\mathcal{T},
\end{equation}
while remaining computationally tractable and efficient in online implementation. For notational brevity, we omit the superscript $0:\mathcal{T}$ when possible. The framework is shown in Figure \ref{frame}.
\begin{remark}
For constraints on samples rather than expectations under distributions, we could formulate the constraints as  $\mathbb{E}_{x^{0:\mathcal{T}} \sim q}[[g(x^{0:\mathcal{T}})]_+] \le 0$ that the probability of any constraint violation is exactly zero under $q$. For the case where the constraints need to be differentiable, we replace with $\mathbb{E}_{x^{0:\mathcal{T}} \sim q}[\frac{g(x^{0:\mathcal{T}})}{1+e^{- g(x^{0:\mathcal{T}})}}] \le 0$ as the approximate estimate of $[g(x^{0:\mathcal{T}})]_+$ .
\end{remark}

This precise formulation highlights the need for a “plug‐and‐play” mechanism to enforce possibly nonconvex constraints (e.g. Discrete Control Barrier Function Constraints in Section \ref{subsec:cbf}) in the reverse diffusion process, motivating our Constrained Diffusers and Online Safety Implementation.

\section{Constrained Diffusers}
\label{headings}
In this section, we present three methods for handling constraints during reverse diffusion sampling: Projected methods, Primal-Dual methods, and Augmented Lagrangian methods. 

\subsection{Projected Diffusion Sampling}
\label{subsec:method_pgd}

This approach leverages Projected Gradient Descent (PGD) to enforce trajectory constraints on samples during the diffusion denoising process. At each denoising step $t$ of the diffusion process, we modify the standard update rule with constraint projection as follows,
\begin{equation}
    x_{t-1} = \Pi_{\mathcal{C}}\left(x_t +\frac{\beta_t}{2} \nabla_{x_t} \log p(x_t) + \sqrt{\beta_t}z \right)
\end{equation}
where $\Pi_{\mathcal{C}}$ denotes the projection operator onto the constraint set $\mathcal{C}$.
The projection operator $\Pi_{\mathcal{C}}$ solves the constrained optimization problem:
\begin{equation}
    \Pi_{\mathcal{C}}(z) = \mathop{\arg\min}_{x \in \mathcal{R}^d} \|x - z\|^2 \quad \text{s.t.} \quad x \in \mathcal{C}
\label{eq:qp}
\end{equation}
Through these updates, the samples will converge to stationary points. Detailed analysis to the statement can be found in Appendix D. This method applies for constraints for samples and the constraints are met at every diffusion step. 
% However, the cost is a significant computational burden. When the constraints become complex or non‑convex, the QP problem \eqref{eq:qp} becomes extremely challenging to solve. Next, we introduce alternative methods to avoid having to solve this QP problem.

\subsection{Primal-Dual Methods (PD)}
\label{subsec:method_pd}

This approach address the constrained optimization problem \eqref{problem} by solving the following problem:
\begin{equation}
    \max_{\lambda \geq 0} \min_q \left\{L(q, \lambda) = \mathrm{KL}(q || p)+\lambda^\mathsf{T} \mathbb{E}_{x \sim q}[g(x)]\right\}. 
\label{eq:lag}
\end{equation}
where $\lambda$ is the corresponding Lagrange multiplier. 
\vspace{1em}
\begin{definition}
A local saddle point of $L(q, \lambda)$ is a point $(q^*, \lambda^*)$ such that for some $r > 0$,  $\forall q \in \mathcal{P}_2(\mathcal{R}^d) \cap \mathcal{B}_{q^*}(r)$ 
and $\forall\lambda \geq 0$, we have
\[
L(q, \lambda^*) \geq L(q^*, \lambda^*) \geq L(q^*, \lambda), 
\]
where $\mathcal{B}_{q^*}(r)$ is a ball centered at $q^*$ with radius $r > 0$ under the 2-Wasserstein metric.
\end{definition}

According to \cite{bookNP, bookPA}, a local saddle point for the maxmin problem is a local optimal solution to the primal constrained optimization problem \eqref{subsec:cbf}. The optimal constrained distribution $q^*$ can be characterized as a tilted version of the original distribution $p(x) \propto e^{-f(x)}$, given by $q^*(x) \propto e^{-(f(x) + \lambda^T g(x))}$. Sampling from $q^*$ can then be achieved by modifying the LMC dynamics to target this tilted distribution. In practice, we solve the maxmin problem by alternatively updating $x$ and $\lambda$: The primal update is to incorporate gradients related to the constraint functions; the dual update is to ascend in $\lambda$ using the gradients of $L(q, \lambda)$ w.r.t. $\lambda$, i.e.,
\begin{equation}
{x_{t-1}} = {x_{t}} + \frac{\beta_t}{2}({\nabla _{x_t}}\log p(x_t)- \lambda_t^T{\nabla _{x_t}} g(x_t)) + \sqrt \beta_t  {{z}_t},
\label{eq:pri}
\end{equation}
\begin{equation}
{{\bf{\lambda }}_{t-1}} = [{{\bf{\lambda }}_{t}} + {\eta _{\bf{\lambda }}}\mathbb{E}_{x \sim q}[g(x_t)]]_+.
\label{eq:dual}
\end{equation}
where $[\cdot]_{+} = \max(0, \cdot)$ denotes projection onto the non-negative orthant, and $\eta_\lambda$ are step sizes. 
\vspace{1em}
\begin{theorem}
Under mild conditions, the sequence of updates \eqref{eq:pri} and \eqref{eq:dual} converges almost surely to a local saddle point, i.e. a local optimal solution to the constrained optimization problem \eqref{problem}, as T goes to infinity.
\end{theorem}

The detailed proof of the theorem can be found in the Appendix E. That is to say, under the proposed update strategy, when a score function already exists, constraint satisfaction can be achieved by their gradients with respect to the samples, eliminating the need for solving a QP.

\subsection{Augmented Lagrangian Methods (ALM)}
\label{subsec:method_al}

This approach enhances constraint handling by introducing a quadratic penalty term into the Lagrangian relaxation which enhances numerical stability\cite{bertsekas2014constrained, li2021augmented}. To handle inequality constraints $\mathbb{E}_{x \sim q}[g(x)] \leq 0$, we introduce slack variables $s \geq 0$ and let $\mathbb{E}_{x \sim q}[g(x)] +s = 0$. Then, we reformulate the constrained optimization problem \eqref{problem} using the augmented Lagrangian:
\begin{equation}
    L_{A}(q, \lambda,s) = \mathrm{KL}(q || p) + \lambda^\mathsf{T} [\mathbb{E}_{x_t \sim q}[g(x_t)] + s] + \frac{\rho}{2}\|\mathbb{E}_{x_t \sim q}[g(x_t)] + s\|^2
    % \max_{\lambda \geq 0} \min_{q,s \geq 0} \left\{L_{AL}(q, \lambda,s) = \mathrm{KL}(q || p) + \lambda^\mathsf{T} [\mathbb{E}_{x_t \sim q}[g(x_t)] + s] + \frac{\rho}{2}\|\mathbb{E}_{x_t \sim q}[g(x_t)] + s\|^2\right\}
\label{eq:auglag_ineq}
\end{equation}
The slack variable $s$ enables exact constraint satisfaction while maintaining differentiability. The update scheme features three components:
\begin{equation}
{x_{t-1}} = {x_{t}} + \frac{\beta_t}{2}\left[\nabla_{x_t}\log p(x_t) - (\lambda_t + \rho_t(\mathbb{E}_{x_t \sim q}[g(x_t)] + s_t))^T\nabla_{x_t} g(x_t)\right] + \sqrt{\beta_t}{z_t},
\label{eq:alm_ineq_pri_x_fo}
\end{equation}
\begin{equation}
s_t = [-\mathbb{E}_{x_t \sim q}[g(x_t)] - \lambda_t/\rho_t]_+,\text{(Slack Update)}
\label{eq:alm_slack}
\end{equation}
\begin{equation}
\lambda_{t-1} = \lambda_t + \rho_t\left(\mathbb{E}_{x_t \sim q}[g(x_t)] + s_t \right),\text{(Dual Update)}
\label{eq:alm_ineq_dual}
\end{equation}

To obtain $\rho \to \infty$, we update the penalty term $\rho_{t-1} = c \cdot \rho_t(c >1)$. These updates will finally converge to the local optimal solution of \eqref{problem} under certain conditions.  Detailed analysis of the statement can be found in Appendix F. 
% \begin{equation}
% {x_{t-1}} = {x_{t}} + \frac{\beta_t}{2}\left[\nabla_{x_t}\log p(x_t) - (\lambda_t + \rho_t(\mathbb{E}_{x_t \sim q}[g(x_t)] + s_t))^T\nabla_{x_t} g(x_t)\right] + \sqrt{\beta_t}{z_t},
% \label{eq:alm_ineq_pri_x_fo}
% \end{equation}
Compared to the primal dual method, this method provides better constraint satisfaction for inequality-constrained diffusion processes.

\section{Online Safety Implementation}
To ensure that the trajectories produced by our model remain safe during online implementation with actions feeding into environments, we incorporate two key components: discrete control barrier functions (DCBFs) and inverse dynamics model based on the following two assumptions: 1) that an inverse dynamic model exists for the systems, 2) that there are no constraints on the inputs.

\subsection{Discrete Control Barrier Function Constraints}
\label{subsec:cbf}

To enhance safety during the implementation stage, we incorporate discrete control barrier functions in our reverse diffusion process. In practice, this condition is typically enforced through a QP at each time step. Since our approach directly generates trajectories, we could directly consider safety constraints between consecutive states $x^\tau$ and $x^{\tau+1}$ at the trajectory level. Specifically, we guarantee that consecutive states in our generated trajectories satisfy the safety condition by rewriting \eqref{dcbf}:
\begin{equation}
h(x^{\tau+1}) \ge (1 - \alpha) h(x^\tau), \quad \forall\, \tau=0,\dots,\mathcal{T},
\end{equation}
here we use a proportional function to replace extended class $\mathcal{K}$ function where $\alpha \in (0, 1]$ is the coefficient that determines how aggressively the system state is required to remain within or approach the safe set. By enforcing this condition directly on the state trajectory, we ensure that the system trajectory evolves safely without explicitly computing control actions during the generation process.

\subsection{Inverse Dynamics Model}
\label{subsec:method_consistency}

When applying the proposed methods at implementation stage in control tasks, the diffuser generates sequences of states. To guarantee the consistency of state and action in the tasks when modifying the reverse diffusion process, we employ an inverse dynamics model (IDM), denoted by  $u^\tau = \text{IDM}(x^\tau, x^{\tau+1})$,
which predicts the action $u$ required to transition from state $x^\tau$ to state $x^{\tau+1}$ after obtaining the constrained state through the proposed algorithms. This ensures that the action executed corresponds to the constraint-satisfying state achieved after the modified reverse diffusion step. The IDM can be learned with deep neural networks when this exists.

The pseudo-code of the complete algorithm is shown below.
\begin{algorithm}[H]
\caption{Constrained Diffusers}
\label{alg:algorithm_label}
    \textbf{Input parameters:}
    Expert data, constraints ${g(x)}$, variance $\mathbf{\beta}$, dual step $\eta_{\bf{\lambda }}$, initial Lagrangian multiplier $\lambda$, initial slack variable $s$, penalty term $\rho$ \\
    1: Use the expert data to train the score function. \\
    2: \textbf{For} $\tau = 0,...,\mathcal{T}$:  // environment timesteps\\
    3: \quad Initialize ${x_T} \sim \mathcal{N}(0, \mathbf{\beta I})$.\\
    4: \quad \textbf{For} $t = T,...,0$:  // diffusion timesteps\\
    5: \quad \quad ${{{x}}_{t-1}} =  $ Constrained Diffuser$\left(x_t, g(\cdot),\beta, \eta_\lambda,\lambda_t,s_t,\rho_t \right)$ // with DCBFs as constraints. \\
    6: \quad \quad // Projected, Primal-Dual, Augmented Lagrangian Methods \\
    7: \quad \textbf{end for} \\
    8: \quad \textbf{return $x_0$ for planning} \\
    9: \quad $u^\tau$ = IDM$(x_0^0,x_0^{1})$ \\
    10:\quad Feed $u^\tau$ into environment. \\
    11: \textbf{end for}
\end{algorithm}

\section{Experiments}
\label{sec:experiments}
In this section, we evaluate our Constrained Diffusers on Maze2D planning tasks, locomotion tasks and pybullet ball running tasks. Our experiments are designed to answer the following key questions:
(1) Can our methods effectively enforce constraints on trajectories generation without retraining?
(2) How does our framework compare to existing techniques in terms of constraint satisfaction, task success, and computational efficiency?
(3) Can our methods adapt to time-varying constraints during deployment?
(4) What are the strengths and weaknesses of the proposed approaches integrated within the diffusion sampling? Detailed experimental settings can be found in Appendix B.

\subsection{Constraint Satisfaction in Maze2D Planning tasks}
\label{subsec:static_constraints}
\begin{figure}[htbp]
  \centering
  \begin{minipage}[b]{0.54\textwidth}
      \centering            
      \subfloat[Maze2d-Umaze]   
      {
          \label{fig:subfig1}\includegraphics[width=0.45\textwidth]{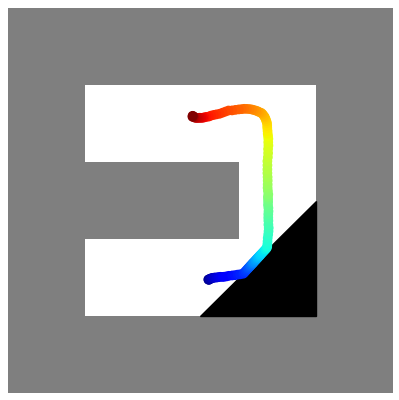}
      }
      \subfloat[Maze2d-Large]
      {
          \label{fig:subfig2}\includegraphics[width=0.45\textwidth]{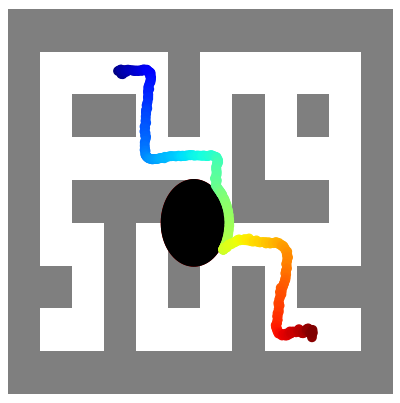}
      }
      \caption{Trajectories generated by Constrained Diffuser in Maze2d Environment with Black obstacles. The blue part shows the start of the trajectories and the red part shows the target. We could find that the trajectories avoid the obstacles and reach the goal}    
      \label{fig:subfig_1}               
  \end{minipage}
  \hfill
  \begin{minipage}[b]{0.44\textwidth}
    \includegraphics[width=\textwidth]{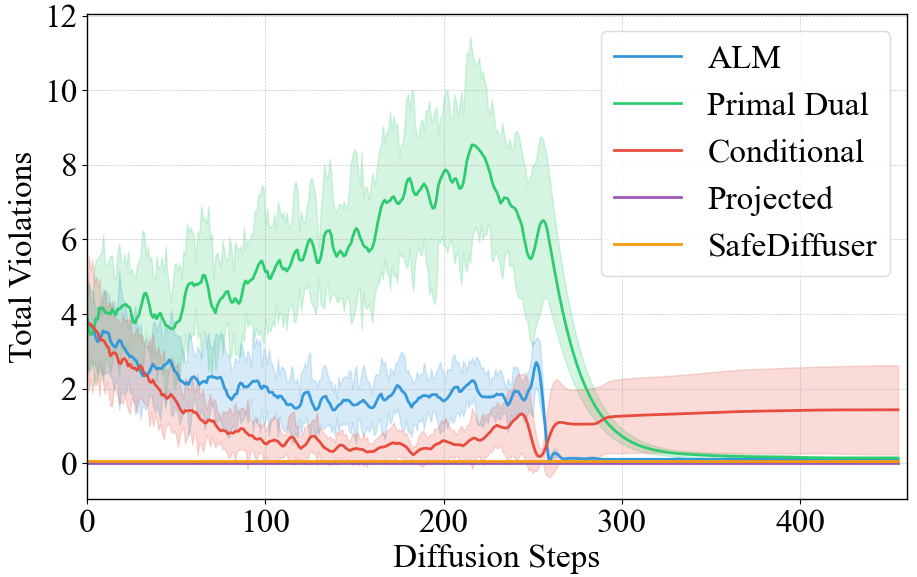}
    \captionof{figure}{Convergence Results for algorithms in Maze2d. Proposed methods can satisfy constraints. ALM shows better convergence and projected methods satisfy constraints throughout reverse diffusion process.}
    \label{fig:new_image}
  \end{minipage}
\end{figure}

First, we evaluate the constrained diffusers on trajectory planning tasks in two Maze2D environments: \textbf{Maze2d-umaze} and \textbf{Maze2d-large}. We set up obstacles in these environments (Figure~\ref{fig:subfig_1}), requiring the agent's planned trajectory to reach the goal while avoiding these obstacles. 
% \vspace{-3mm}
\begin{table}
  \caption{Performance comparison for Maze2D tasks. We can find that the our algorithm could almost satisfy constraints and reduce computational time, compared to baseline algorithms. We report the mean and the standard error over 10 random environments.}
  \label{table_static}
  \centering
  \begin{tabular}{lllll}
    \toprule 
    \textbf{Env} & \textbf{Algorithm} & \textbf{Distance} & \textbf{Total Violations} & \textbf{Time/s} \\
    \midrule
    \multirow{6}{*}{maze2d-umaze} 
    & Diffuser & 0.00 & 7.316 $\pm$ 0.884 & 0.0016 \\
    & Conditional & 23.96 $\pm$ 2.14 & 1.109 $\pm$ 0.834 & 0.0017 \\
    & SafeDiffuser & 42.81 $\pm$ 8.66 & \textbf{0.000 $\pm$ 0.000} & 0.1208 \\
    & Projected & 10.24 $\pm$ 1.39 & \textbf{0.000 $\pm$ 0.000} & 0.0107 \\
    & Primal Dual & 9.78 $\pm$ 1.14 & 0.013 $\pm$ 0.032 & \textbf{0.0017} \\
    & ALM & 9.81 $\pm$ 1.65 & 0.002 $\pm$ 0.002 & 0.0018 \\
    \midrule  % 
    \multirow{6}{*}{maze2d-large}  %
    & Diffuser & 0.00 & 12.400 $\pm$ 2.440 & 0.0022 \\
    & Conditional & 194.31 $\pm$ 13.25 & 1.653 $\pm$ 1.124 & 0.0027 \\
    & SafeDiffuser & 196.25 $\pm$ 33.06 & \textbf{0.000 $\pm$ 0.000} & 0.2219 \\
    & Projected & 199.95 $\pm$ 37.45 & \textbf{0.000 $\pm$ 0.000} & 0.0083 \\
    & Primal Dual & 187.52 $\pm$ 33.80 & 0.069 $\pm$ 0.056 & 0.0028 \\
    & ALM & 195.85 $\pm$ 21.82 & 0.033 $\pm$ 0.044 & 0.0029 \\
    \bottomrule
  \end{tabular}
\end{table}
These obstacles can be expressed using linear and quadratic inequalities, respectively. We explicitly incorporate these constraints into our trajectory generation process. We compare our algorithms against three baselines: Diffuser (basic trajectory generation via diffusion), Conditional Diffuser\cite{ajay2022conditional} (constraint-conditioned diffusion for trajectory generation), and SafeDiffuser\cite{xiao2023safediffuser} (incorporating CBFs for denoising process). Table \ref{table_static} illustrates the Euclidean distance from the original trajectory, total constraint violations throughout the whole planned trajectories $\Sigma_{\tau=0}^\mathcal{T}[g(x^\tau)]_+$, and computation time per diffusion step for each algorithm in both environments.

From Table \ref{table_static}, conditional diffusion struggles to handle constraints in environments. SafeDiffuser and Projected method fully satisfies the constraint at the cost of much computational time. The other two methods almost satisfy the constraints while maintaining computational efficiency and ALM handle constraints better. This demonstrates the ability of our proposed approach: being nearly consistent with expert behavior while flexibly handling constraints to ensure safety, all without retraining.

\subsection{Safe control in locomotion tasks}
% \label{subsec:cbf_analysis} \vspace{-3mm}
% \begin{wrapfigure}[22]{r}{0.3\textwidth} 
%   \centering
%   \begin{minipage}[b]{0.8\linewidth} 
%     \centering
%     \includegraphics[width=\linewidth]{hopper_1.png}
%     \caption{Hopper}
%     \label{fig:a}
%   \end{minipage}
%   \vspace{0.2cm} 
%   \begin{minipage}[b]{0.8\linewidth} 
%     \centering
%     \includegraphics[width=\linewidth]{swimmer_1.png}
%     \caption{Swimmer}
%     \label{fig:b}
%   \end{minipage}
% \end{wrapfigure}
% \begin{wrapfigure}{r}{0.32\textwidth} 
%   \centering
%   \includegraphics[width=0.3\textwidth]{hopper_1.png}
%   \caption{Hopper in Gymnasium}
%   \label{fig:example}
% \end{wrapfigure}
Next, we test the performance of our algorithm on planning and control tasks in two Gymnasium MuJoCo environments, \textbf{Hopper} and \textbf{Swimmer} \cite{towers2024gymnasium}. For Hopper, we constrain the angular velocity of the thigh hinge within a threshold; for Swimmer, we require the angular velocity of the second rotor to remain within a safe range. In these tasks, we employ DCBF as constraints and used IDM to generate actions. We selected Diffuser, Conditional Diffuser, and SafeDiffuser as baseline methods to compare. The metrics include total constraint violations that the amount of the violations for the planning trajectory generated by diffusion, rewards, constraint violations and violation rates when interacting with environments through the actions from IDM, and computation time per diffusion step. The results in summarized in Table \ref{table_2}. SafeDiffuser methods applies only to the planning stage so we do not compare their metrics in implementation stage. Our algorithms have better performance in both planning and implementation stage. We also show that DCBFs are effective in Appendix C.

\begin{table}[H]
  \caption{Performance comparison for locomotion tasks.  Projected method has the best results in planning stage constraint satisfaction, while the other two methods trade off computational efficiency without compromising safety. We report the mean and standard error over 10 random environments.}
  \label{table_2}
  \centering
  \begin{tabular}{lllllll}
    \toprule 
    \textbf{Env} & \textbf{Algorithm} & \textbf{Reward} & \makecell{\textbf{Planning} \\ \textbf{Violations}} & \makecell{\textbf{Impl.} \\ \textbf{Violations}} & \makecell{\textbf{Violation} \\ \textbf{Rates(\%)}} & \textbf{Time/s} \\
    \midrule
    \multirow{6}{*}{Hopper} 
    & Diffuser & 3592 $\pm$ 37 & 2.55 $\pm$ 0.21 & 2.61 $\pm$ 0.21 & 4.82 $\pm$ 0.60 & 0.0026  \\
    & Conditional & 3608 $\pm$ 36 & 1.06 $\pm$ 0.19 & 1.38 $\pm$ 0.27 & 4.74 $\pm$ 0.83 & 0.0027  \\
    & SafeDiffuser & - & \textbf{0.00 $\pm$ 0.00} & - & - & 0.9307  \\
    & Projected & 3547 $\pm$ 13 & \textbf{0.00 $\pm$ 0.00} & \textbf{0.29 $\pm$ 0.09} & 0.29 $\pm$ 0.08 & 0.9305  \\
    & Primal Dual & 3551 $\pm$ 10 & 0.10 $\pm$ 0.03 & 0.43 $\pm$ 0.12 & 0.41 $\pm$ 0.30 & 0.0027  \\
    & ALM & 3553 $\pm$ 14 & 0.08 $\pm$ 0.02 & 0.38 $\pm$ 0.16 & \textbf{0.09 $\pm$ 0.11} & 0.0031  \\
    \midrule
    \multirow{6}{*}{Swimmer} 
    & Diffuser & 57.4 $\pm$ 8.7 & 8.38 $\pm$ 0.87 & 1.12 $\pm$ 0.26 & 1.30 $\pm$ 0.26 & 0.0026  \\
    & Conditional & 58.8 $\pm$ 5.6 & 8.02 $\pm$ 0.78 & 1.00 $\pm$ 0.31 & 1.11 $\pm$ 0.43 & 0.0027  \\
    & SafeDiffuser & - & \textbf{0.00 $\pm$ 0.00} & - & - & 0.9301  \\
    & Projected & 58.7 $\pm$ 4.1 & \textbf{0.00 $\pm$ 0.00} & 0.61 $\pm$ 0.25 & 0.81 $\pm$ 0.26 & 0.9049  \\
    & Primal Dual & 86.4 $\pm$ 9.2 & 0.17 $\pm$ 0.08 & 0.02 $\pm$ 0.02 & 0.07 $\pm$ 0.07 & \textbf{0.0027}  \\
    & ALM & 88.7 $\pm$ 5.4 & 0.08 $\pm$ 0.05 & \textbf{0.00 $\pm$ 0.02} & \textbf{0.03 $\pm$ 0.04} & 0.0029  \\
    \bottomrule
  \end{tabular}
\end{table}

\subsection{Adaptability to Time-varying Constraints}
\begin{wrapfigure}[13]{r}{0.32\textwidth} 
  \centering  \includegraphics[width=0.30\textwidth]{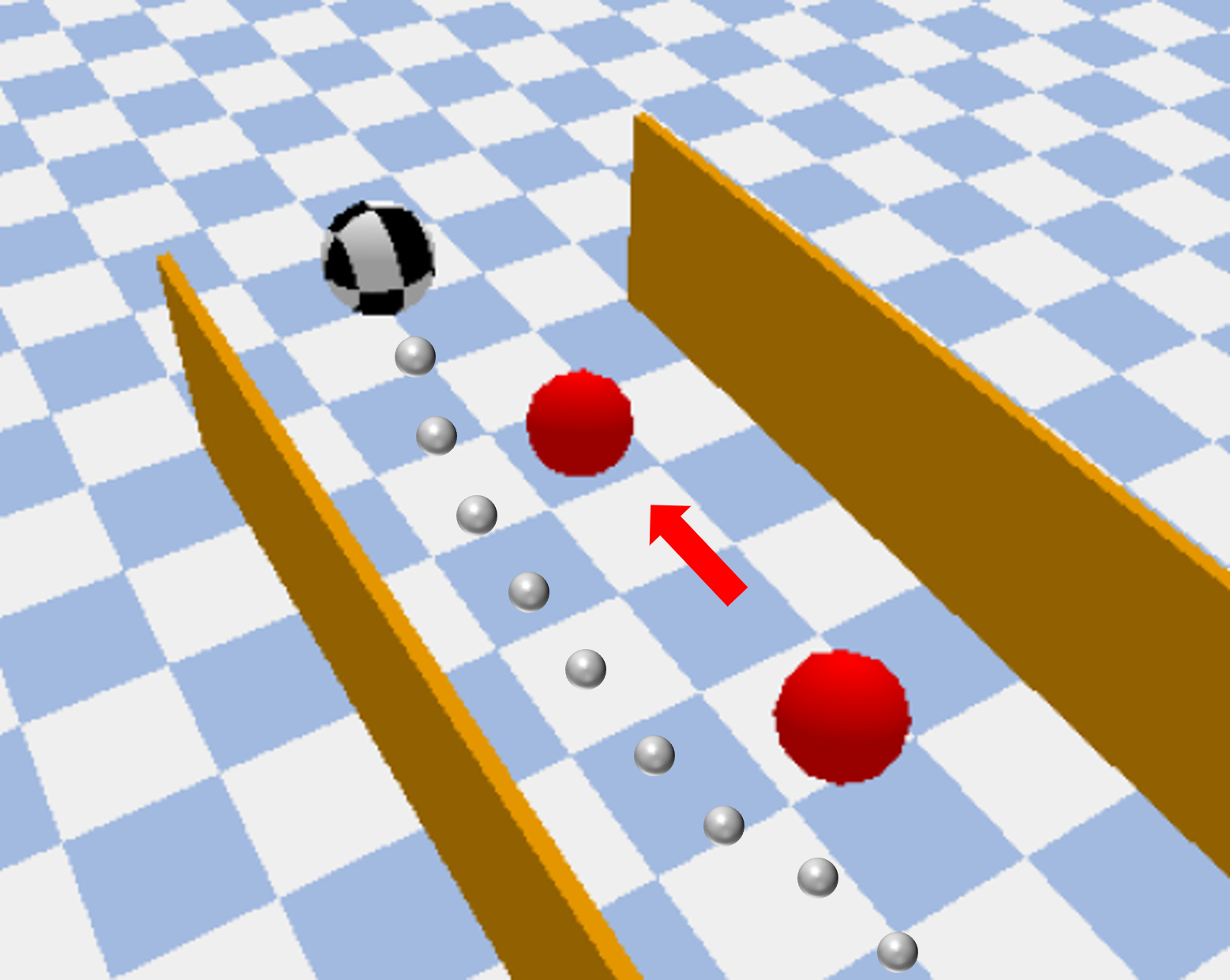}
  \caption{PybulletBallRunning tasks with the black-white ball and a red moving obstacles ball.}
  \label{running}
\end{wrapfigure}Finally, we evaluate the adaptability of our proposed algorithm in the PyBullet SafetyBallRun environment, where safety constraints varies over time. In this setup, we have a obstacle ball that moves in the space, requiring the ball to avoid collision while continuing to move. We assume that the agent ball have no knowledge of the obstacle's dynamics and can only observe its current states, making planning infeasible. During implementation, we construct the discrete control barrier functions (DCBF) based on the current observation of obstacles as the constraints in the constrained diffuser to generate a trajectory. We use Diffuser and Conditional Diffuser as baseline algorithms. We compare rewards, constraint violations and violation rates during implementation stage, and computation time per diffusion step in the experiment. The results are summarized in Table \ref{table_dyn}.

\begin{table}[H]
  \caption{Performance comparison for pybullet ball running tasks. The results indicate our methods effectively adapt to time-varying constraints, with low violation rates. We use closed-form solution in projected mathods. We report the mean and the standard error over 10 random environments.}
  \label{table_dyn}
  \centering
  \begin{tabular}{lccccc}
    \toprule 
    \textbf{Env} & \textbf{Algorithm} & \textbf{Reward} & \makecell{\textbf{Total Impl.} \\ \textbf{Violations}} & \makecell{\textbf{Violation} \\ \textbf{Rates(\%)}} & \textbf{Time/s} \\
    \midrule
    \multirow{5}{*}{SafetyBallRun} 
    & Diffuser & 646.4 $\pm$ 30.3 & 2.20 $\pm$ 0.77  & 37.7 $\pm$ 5.1 & 9.6e-3  \\
    & Conditional & 616.5 $\pm$ 33.9 & 1.55 $\pm$ 0.53 & 33.1 $\pm$ 11.1  & 9.7e-3  \\
    & Projected & 586.7 $\pm$ 33.7 & 1.01 $\pm$ 0.63 & 25.8 $\pm$ 13.9 & 2.6e-2  \\
    & Primal Dual & 722.2 $\pm$ 118.6 & 0.02 $\pm$ 0.03  & 2.2 $\pm$ 3.2 & 1.3e-2  \\
    & ALM & 696.1 $\pm$ 99.9 & \textbf{0.02 $\pm$ 0.02} & \textbf{1.6 $\pm$ 3.4} & 2.1e-2  \\
    \bottomrule
  \end{tabular}
\end{table}
Compared to the baselines, we can ensure effective collision reduction with less decision time. In addition, we also observed an improvement in the reward during implementation. We interpret this as the adjustment bringing the ball closer to the target.

\section{Related Work}

\noindent \textbf{Diffusion Models and Policy Representation}  Diffusion models \cite{sohl2015deep,ho2020denoising} have demonstrated significant potential in image\cite{dhariwal2021diffusion, ho2022classifier} and text generation\cite{nichol2022glide, saharia2022photorealistic}. It gradually adds noise to the data, transforming it into a random distribution over time \cite{song2019generative, song2020score} and learns to iteratively denoise the data by reversing the forward diffusion steps \cite{anderson1982reverse}. This success extended to planning tasks \cite{janner2022planning, ajay2022conditional}, outperforming traditional imitation learning methods through the ability to model complex distributions \cite{chi2023diffusion,ren2024diffusion}. Recent works focus on trajectory optimization through diffusion models \cite{pan2024model, li2024diffusolve} while they require interactions with environments in the optimization process. Some researchers have also investigated how diffusion can be combined with reinforcement learning to guide strategy improvement\cite{zheng2024safe}. However, they didn't handle constraints in the trajectory generation and online implementation through diffusion.

\noindent \textbf{Constrained Sampling} 
 Constrained sampling methods have been developed for support constraints in the standard sampling process \cite{fishman2023diffusion}. Projected Langevin Monte Carlo extends the algorithm to compactly supported measures through a projection step when sampling from a log-concave distribution \cite{bubeck2018sampling}. Primal-dual methods simultaneously sample from the target distribution and constrain it through gradient descent-ascent dynamics in the Wasserstein space \cite{chamon2024constrained, liu2021sampling}. In addition, some Langevin Monte Carlo methods based on mirror maps \cite{ahn2021efficient}, barriers~\cite{kook2022sampling}, and penalties~\cite{gurbuzbalaban2022penalized} have also demonstrated significant advantages in constrained sampling. However, they didn't bridge the connections with diffusion models and didn't extend them to nonconvex settings. 

\noindent \textbf{Safe Control Policy} 
 Safe control has emerged as a critical research area in robotics and autonomous systems\cite{brunke2022safe}.  Traditional model-based approaches, such as Hamilton-Jacobi reachability analysis\cite{mitchell2007comparing} and Model Predictive Control \cite{garcia1989model}, offer rigorous safe guarantees. Reinforcement learning algorithms addressing safety concerns have also attracted considerable attention \cite{garcia2015comprehensive, achiam2017constrained, ray2019benchmarking}. Additionally, safe exploration techniques incorporate safety layers to adjust actions \cite{amos2017optnet}, while other algorithms improve policy robustness against potential safety risks \cite{JMLR:v18:15-636}. Control barrier functions (CBFs) have been widely applied in the field of safe control \cite{ames2019control, agrawal2017discrete}. Recent work employ CBFs in diffusion to ensure that generated trajectories avoid obstacles \cite{xiao2023safediffuser}. Yet, the need to solve a quadratic programming problem at each diffusion step results in high computational complexity. 

\section{Conclusions}
\label{others}

In this work, we proposed constrained diffusers in planning and control tasks. The model integrates constraints directly into pre-trained diffusion models without requiring retraining or architectural changes. By re-intepreting safe trajectory generation as a constrained sampling problem, we introduce three methods--Projected, Primal-Dual and Augmented Lagrangian methods--to realize the constrained reverse Langevin sampling. Then we introduce DCBFs as constraints in the Constrained Diffusers and use inverse dynamics models to obtain actions, ensuring safety in the implementation stage. We demonstrated them in Maze2D, robotic locomotion, and pybullet ball running tasks demonstrate that the proposed constrained diffusion model not only meets safety requirements but also reduces computational time compared to existing approaches.

% \begin{ack}
% Use unnumbered first level headings for the acknowledgments. All acknowledgments

% Do {\bf not} include this section in the anonymized submission, only in the final paper. You can use the \texttt{ack} environment provided in the style file to automatically hide this section in the anonymized submission.
% \end{ack}

% \section*{References}

% References follow the acknowledgments in the camera-ready paper. Use unnumbered first-level heading for
% the references. Any choice of citation style is acceptable as long as you are
% consistent. It is permissible to reduce the font size to \verb+small+ (9 point)
% when listing the references.
% Note that the Reference section does not count towards the page limit.
% \medskip

{
\small

% [1] Alexander, J.A.\ \& Mozer, M.C.\ (1995) Template-based algorithms for
% connectionist rule extraction. In G.\ Tesauro, D.S.\ Touretzky and T.K.\ Leen
% (eds.), {\it Advances in Neural Information Processing Systems 7},
% pp.\ 609--616. Cambridge, MA: MIT Press.

% [2] Bower, J.M.\ \& Beeman, D.\ (1995) {\it The Book of GENESIS: Exploring
%   Realistic Neural Models with the GEneral NEural SImulation System.}  New York:
% TELOS/Springer--Verlag.

% [3] Hasselmo, M.E., Schnell, E.\ \& Barkai, E.\ (1995) Dynamics of learning and
% recall at excitatory recurrent synapses and cholinergic modulation in rat
% hippocampal region CA3. {\it Journal of Neuroscience} {\bf 15}(7):5249-5262.

% [4] M. Janner, Y. Du, J. Tenenbaum, and S. Levine, “Planning with Diffusion for Flexible Behavior Synthesis,” in Proceedings of the 39th International Conference on Machine Learning, PMLR, Jun. 2022, pp. 9902–9915.

\bibliographystyle{unsrt}
\bibliography{ref}
}

\end{document}